# Characterization of TOF-PET Detectors Based on Monolithic Blocks and ASIC-Readout

Efthymios Lamprou, Andrea González-Montoro, Gabriel Cañizares, Victor Ilisie, John Barrio,
Filomeno Sánchez, Antonio J. González, José Maria Benlloch

*Abstract*— **The aim of this work is to show the potential capabilities of monolithic crystals coupled to large SiPM arrays, to be considered as detector blocks for PET scanners enabling Time Of Flight (TOF) capabilities. Monolithic blocks allow one to decode the 3D photon impact position. This approach, along with TOF information, can be of high interest in clinical Positron emission tomography (PET) applications where a typical ring configuration is used. In this manuscript, we evaluate an ASIC-based readout for digitizing all signals coming from analog photosensors. Validation results with one-to-one coupling resulted in a Coincidence Time Resolution (CTR) of 202 ps FWHM.**

**Providing timing resolution when using detectors based on monolithic crystals is however challenging. The wide distribution of scintillation light on the photosensors causes a poor SNR, which makes the system sensible to false triggering and to time walk errors. In this direction, we present a calibration method, designed to correct all recorded timestamps and also to compensate variations in time-paths among all channels. Thereafter, a CTR improvement nearing 45% is observed for all measurements. Moreover, we show a novel approach that describes the use of averaging methods to assign a single timestamp to each gamma impact. This approach results in a further improvement of the CTR in the range of 100 ps FWHM, reaching a time resolution of 585 ps FWHM when using a large 50×50×10 mm³ LYSO scintillator coupled to an 8×8 SiPM (6×6 mm²) array. These pilot studies show detector capabilities regarding TOF information when using monolithic scintillators.**

*Index Terms*- ASIC, Dedicated-PET, Monolithic blocks, SiPMs, TOF-PET

## I. Introduction

RECENT advances in front-end electronics have made feasible the development of high performance gamma ray detectors for Positron Emission Tomography (PET) [1]. Nowadays, PET systems providing accurate timing resolution are being established in the clinical practice [2],[3]. Typically, Time Of Flight (TOF)-PET detectors are composed by analog photosensors and the One-To-One coupling methods (crystal-to-photosensor element). Even if this approach is rather optimal in terms of Coincidence Time Resolution (CTR), it lacks of performance when comes to decode the photon Depth Of Interaction (DOI) [4].

A gamma ray detector, capable of resolving gamma-ray impacts in all 3D coordinates as well as in time, could be of high interest in multiple applications as it will offer the possibility for more accurate determination of the positron-electron annihilation point inside the subject of study, leading to a significant reduction of both noise and diagnostic error.

When using monolithic blocks, the wide spread of scintillation light results on a limited Signal-To-Noise ratio (SNR) comparing to the One-To-One coupling. This eventually leads to false triggering by dark counts from Silicon Photomultiplier (SiPM) and to timestamps uncertainty due to signals jitters and time walk errors [5].

In this work, we are evaluating PET detectors based on monolithic blocks and analog SiPM arrays for their potential integration in a clinical TOF-PET system. Aiming to compensate these challenges, we have designed a calibration method to correct each timestamp measured for its uncertainty, achieving also a time resolution homogeneity among all channels. All these methods are being described in the present report. Several detector configurations are being evaluated in terms of timing, energy and position accuracies.

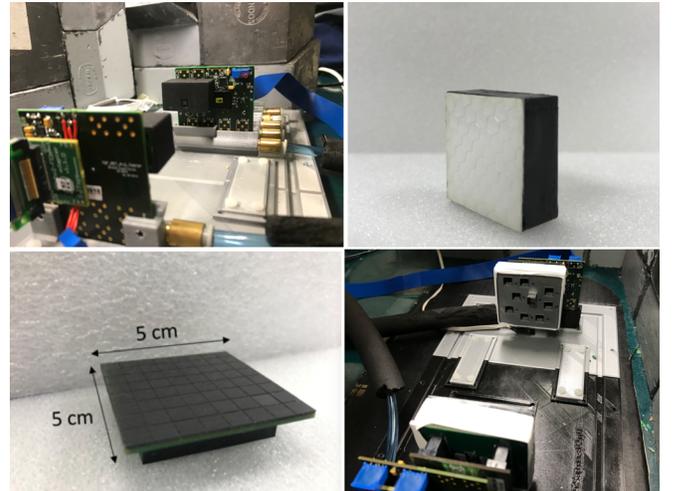

Fig. 1. Top: Left. Experimental Set-Up based on the TOFPET2 ASIC and the PM3325-WB SiPM arrays. Right. Monolithic block, black painted with the entrance layer coupled to a RR layer. Bottom: Left. A large 8×8 J-Series SiPM photosensor. Right. Experimental Set-Up based on a monolithic block and a single LYSO pixel for the time alignment calibration.

Manuscript submitted June 2, 2018. This project has received funding from the European Research Council (ERC) under the European Union's Horizon 2020 research and innovation program (grant agreement No 695536). It has also been supported by the Spanish Ministerio de Economía, Industria y Competitividad under Grant TEC2016-79884-C2-1-R. E. Lamprou, A. Aguilar, A. González-Montoro, G. Cañizares, V. Ilisie, J. Barrio, F. Sánchez, A.J. González and J.M. Benlloch are with the Institute for Instrumentation in Molecular Imaging, i3M CSIC-UPV, Valencia, Spain (e- mail: e.lamprou@i3m.upv.es).

## II. MATERIALS AND METHODS

In order to independently process each SiPM photosensor element, a method that seems the optimal when aiming to accurate TOF, we make use of a multichannel ASIC-based readout. In detail, we have used the TOFPET2 ASIC (PETsys, Lisbon, Portugal), a 64-channel chip that integrates quad-buffered TDCs (30 ps time binning) and linear charge integrators for each channel [6]. Its low configurable threshold for timing and its high event rates capabilities, makes it suitable for light sharing applications.

Aiming to characterize the ASIC performance and reveal the limits in terms of CTR, two small LYSO crystals of size $3 \times 3 \times 5$ mm$^3$ were coupled to the KETEK (Munich, Germany) PM3325-WB SiPM (Fig. 1 top-left) and coincidences measurements were carried out.

A pair of large $8 \times 8$ SensL (Cork, Ireland) J-Series SiPM arrays with $6 \times 6$ mm$^2$ active area each were coupled to LYSO crystal arrays of $32 \times 32$ crystal elements (1.6 mm pixel size, 6 mm height) for a first evaluation of spatial and energy resolution, already enabling light sharing mechanisms. Finally, the aforementioned SiPM arrays were used to characterize a monolithic LYSO block of $50 \times 50 \times 15$ mm$^3$ with black painted lateral walls and a retro-reflector (RR) layer [7] coupled to the entrance face of the crystal. The raw time resolution recorded with the monolithic based detector required a calibration procedure (Fig. 1 bottom-right). The method implies the generation of a Look-Up-Table (LUT) that contains an offset value for all timestamps depending on the ASIC channel and impact energy, and it is used in all following measurements for correcting the recorded timestamps. Further details are provided in the following lines.

## III. RESULTS

Results obtained with the PM3325 SiPM pixels and the small crystals verified the ASIC performance. A CTR of 202 ps FWHM was obtained with a good energy resolution (Fig. 2) of 8% FWHM. The measurements temperature was stabilized to 18 ºC to avoid significant drifts in gain. Moreover, SiPM bias and threshold scans did not reveal further improvements.

Following the initial CTR evaluation, 2D images were obtained using the large $8 \times 8$ SensL arrays. As it can be seen in Fig. 3 left, we have been able to clearly resolve all $32 \times 32$ crystal pixels along with a good spatial resolution and SNR despite the big active area of the photosensors and the small crystal size.

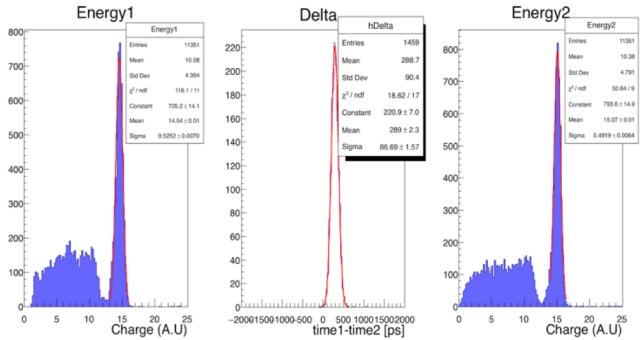

Fig. 2.: Time resolution determined at 202 ps FWHM using PM3325-WB SiPMs and one-to-one-coupling. Temperature was stabilized at 18˚C.

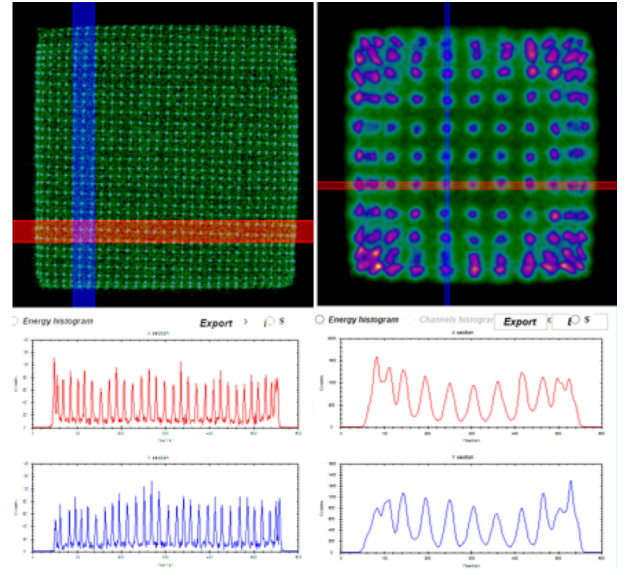

Fig. 3: Left. Flood map showing all 32 x 32 crystal elements of 1.6 mm size, along with a good SNR. Right. A flood map of the monolithic based detector with a collimated source array attached.

An estimation for the spatial resolution was also obtained for the case using monolithic based detectors. A source array ($11 \times 11$ $^{22}$Na, 1 mm diameter, 4.6 mm pitch) was attached to one detector and a spatial resolution of 1.8 mm FWHM was estimated for the central region, slightly worsening towards the crystal edges (see Fig. 3 right).

### A. Timing resolution characterization with monolithic blocks

As mentioned above, the wide scintillation light distribution results in a poor SNR. As more as 20 SiPMs are typically fired per gamma impact when using monolithic blocks. Thus, when using analog SiPM photosensors intense degradation in time resolution occurs, as the system becomes sensitive to false triggering by the dark counts and to time walk error. To initially characterize the CTR, we carried out coincidence measurements with a reference detector composed by a single LYSO pixel $6 \times 6 \times 15$ mm$^3$. The analysis of the raw data revealed a CTR significantly worse than those obtained when using smaller crystals and one-to-one coupling, as expected. A time resolution of 1.5 ns FWHM was found.

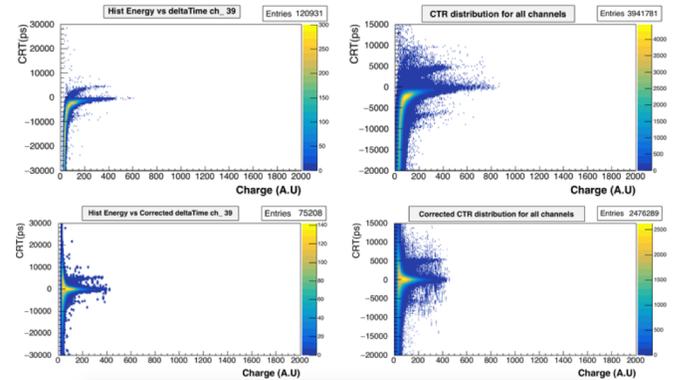

Fig. 4. Top: 2D histograms of the CTR measured as a function of the energy for three pairs of channels before calibration. Bottom Left, 2D histograms of the corrected CTR as a function of the energy for a pair of channels after calibration. Bottom Right: CTR distribution as a function of the energy for all ASIC channels after correcting each timestamp measured.

First step in improving the CTR is to carry out a calibration method that is designed to correct each recorded timestamp, by firstly compensating the time walk error. We plotted all the time differences recorded between detectors, for each pair of channels, as a function of the energy of each impact in 2D contour plots. A filter in the energy of the photopeak of the reference detector was also applied. As it is clearly shown in Fig. 4 top, CTR worsens for low energy channels, as the influence by the time walk is getting more intense. By projecting the 2D histogram in small energy steps and by fitting these projections to Gaussian distributions we were able to build a reference LUT that contains the Gaussian centroid and sigma for all the energy range. Using this information, we eventually were able to correct each timestamp by summing the Gaussian Centroid to the timestamp recorded as an offset. Eventually, after correcting all timestamps belonging to small energy ranges, the distributions are centered to zero (Fig. 4 bottom). By applying the same approach in all channels we were able to compensate the time walk error as well as to calibrate the timepaths' misalignment among the SiPMs and ASIC channels. Eventually, a CTR homogeneity among all channels was achieved, improving significantly by a factor of 45% the time resolution in all following measurements. (Fig. 4 bottom right).

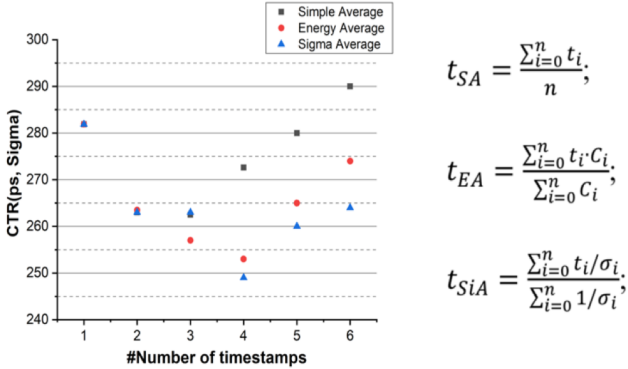

Fig. 5.: Plots showing the time resolution measured as a function of the number of averaging timestamps for three methods: simple average, average weighted by energy, and average weighted by sigma.

Approaches for averaging timestamps of several SiPMs fired for determining with more precision the time of interaction were tested. Three methods, namely simple average, average weighted by energy, and average weighted by the sigma recorded during the calibration process, were implemented. The analysis showed that all three methods up to a number of timestamps, contribute to an improvement of the time resolution (Fig. 5). The sigma averaging method of four timestamps though, seems to provide the best results. The time resolution presented in Figure 6 shows that we can eventually reach a CTR of 585 ps FWHM, which improves by a factor of more than 2 the CTR obtained from the raw timestamps data (1.5 ns).

Finally, aiming to validate the CTR values presented in this work, we recorded the centroids of CTR distributions when moving the source in steps of 2.5 cm between the monolithic detector and the reference one. By plotting the recorded values as a function of the theoretical values we show that the system timing linearity was improved after correcting each timestamp and after using the averaging methods (see Fig. 7). This confirms that our suggested methods can significantly decrease the time uncertainty introduced by time walk error, signal jitter and different signal paths to the photosensors, thus providing more reliable results.

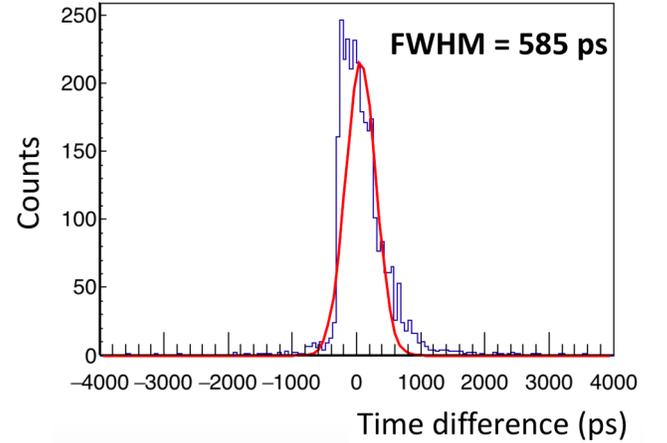

Fig. 6.: Time differences measured for the monolithic based detector after applying the timestamp correction and a sigma weighted averaging method of 4 timestamps. A CTR of 585 ps FWHM is obtained.

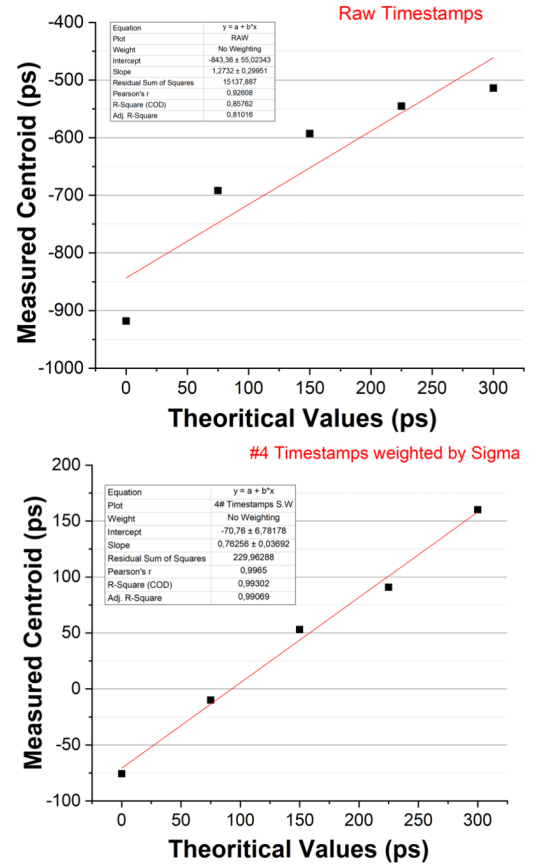

Fig. 7.: Top. Linearity of the Gaussian Centroids recorded using raw timestamps. Bottom. Linearity of the Gaussian Centroids after applying the Time-Alignment calibration and the sigma weighted averaging method of 4 timestamps.

## IV. Conclusion

To our knowledge, this work is novel with regards to the accurate determination of TOF information from detectors based on monolithic blocks and analog SiPMs using TOFPET2 ASIC. Several studies have been presented so far where monolithic blocks are read out using digital SiPMs reaching sub-200 ps FHWM [8]. Those photosensors exhibit the capability for disabling the noisiest microcells facilitating the extrapolation of TOF information.

We present methods to deal with the challenges arising when combining analog SiPMs and monolithic blocks. The time alignment calibration presented above, as proven, returns an effective method to correct the uncertainty on low gain signals and can be applied in configurations using light-sharing approaches. Additionally, the timestamps averaging method allowed us to further improve the CTR reaching eventually a time resolution of 580 ps FWHM for a thick monolithic block, improving by a factor of 2 the CTR obtained from the recorded raw timestamps.

Several steps remain to be done towards the aim of 300 ps FWHM between large monolithic detector blocks. Correction for the time propagation of the photons based on the DOI information is being currently studied, which for thick monolithic crystals, can be critical [9]. SiPM bias and threshold scans need also to be carried out, in order to find out if a more optimal system configuration is possible.

## References

[1] P. Lecoq, "Development of new scintillators for medical applications," Nucl. Instrum. Meth. A 809, 130, 2016.

[2] S. Vandenberghe et al., "Recent developments in time-of-flight PET," EJNMMI Phys 3, 3, 2016.

[3] S. Surti and J.S. Karp, "Advances in time-of-flight PET," Phys. Med. 32(1): 12–22, 2016.

[4] A. J. González et al., "A PET Design Based on SiPM and Monolithic LYSO Crystals: Performance Evaluation," IEEE Trans. Nucl. Sci. 63, 2471, 2016.

[5] E. Lamprou, et al., "PET detector block with accurate 4D capabilities," Nucl. Inst. Meth. A, in press, 2018. DOI: https://doi.org/10.1016/j.nima.2017.11.002.

[6] T. Niknejad, et al., "TOFPET2: a high-performance ASIC for time and amplitude measurements of SiPM signals in time-of-flight applications," Journal of Inst. 11, 2016.

[7] A. Gonzalez-Montoro, et al., "Performance Study of a Large Monolithic LYSO Detector with Accurate Photon DOI Using Retroreflector Layers," IEEE Trans. Rad. Plasma Med. Scie. 1, 229-237, 2017.

[8] H.T. van Dam, et al., "Sub-200 ps crt in monolithic scintillator pet detectors using digital sipm arrays and maximum likelihood interaction time estimation," Phys. Med. Biol. 58, 3243–3257, 2013.

[9] K. Shibuya et al., "Timing resolution improved by DOI information in an LYSO TOF-PET detector," 2007 IEEE NSS Conference Record, Honolulu, HI, 3678-3680, 2007.